\documentclass[10pt,twocolumn,nocopyrightspace]{sigplanconf-eurosys}

\usepackage{amsmath, amsthm, amsfonts, amssymb}
\usepackage{graphicx}
\usepackage{fancyvrb}
\usepackage{listings}
\usepackage{color}
\usepackage{multirow}
\usepackage{url}
\usepackage[font={small}]{caption}
\usepackage{subfig}
\usepackage{xspace}
\usepackage[ruled,vlined]{algorithm2e}
\usepackage{enumitem}

\definecolor{stcolor}   {rgb}{.0,.0,.0}
\definecolor{kwcolor}   {rgb}{.1,.1,.6}
\definecolor{typecolor} {rgb}{.2,.2,.9}
\definecolor{funccolor} {rgb}{.5,.2,.2}
\definecolor{varcolor}  {rgb}{.2,.2,.5}
\definecolor{cmtcolor}  {rgb}{.2,.5,.2}
\definecolor{constcolor}{rgb}{.7,.2,.7}
\definecolor{backcolor} {rgb}{.8,.8,.8}

\newcommand{\ignore}[1]{}

\newcommand{\tf}{\textsc{TensorFlow}\xspace}
\newcommand{\mxnet}{\textsc{MXNet}\xspace}
\newcommand{\name}{\textsc{SoyBean}\xspace}
\newcommand{\schR}{\texttt{R}}
\newcommand{\schC}{\texttt{C}}
\newcommand{\schr}{\texttt{r}}
\newcommand{\schred}{\texttt{red}}
\newcommand{\schP}{\texttt{P}}

\DeclareMathOperator*{\argmin}{arg\,min}
\newcommand{\mytilde}{\raise.17ex\hbox{$\scriptstyle\mathtt{\sim}$}}

\newtheorem{defi}{Definition}
\newtheorem{thm}{Theorem}

\setlist[itemize]{noitemsep}

\begin{document}






\title{Unifying Data, Model and Hybrid Parallelism in Deep Learning via Tensor Tiling}

           
\authorinfo{Minjie Wang\and Chien-chin Huang\and Jinyang Li}
           {New York University}
           {}

\maketitle

%
%

\begin{abstract}
	Deep learning systems have become vital tools across many fields, but the
	increasing model sizes mean that training must be accelerated to maintain such
	systems' utility. Current systems like \tf and \mxnet focus on one
	specific parallelization strategy, data parallelism, which requires large
	training batch sizes in order to scale.  We cast the problem of finding the
	best parallelization strategy as the problem of finding the best tiling to
	partition tensors with the least overall communication.  We propose an
	algorithm that can find the optimal tiling.  Our resulting parallelization
	solution is a hybrid of data parallelism and model parallelism.  We build the
	\name system that performs automatic parallelization. \name automatically
	transforms a serial dataflow graph captured by an existing deep learning system
	frontend into a parallel dataflow graph based on the optimal tiling it has
	found.  Our evaluations show that \name is $1.5\times-4\times$ faster than pure
	data parallelism for AlexNet and VGG. We present this automatic tiling in a new
	system, \name, that can act as a backend for \tf, \mxnet, and others.
\end{abstract}

\section{Introduction}
\label{sec:intro}
Deep neural networks (DNNs) have delivered tremendous improvements across many
machine learning tasks, ranging from computer vision~\cite{krizhevsky2012imagenet,vggnet}
and speech recognition~\cite{neuralspeech,speech:hinton} to natural
language processing~\cite{seqtranslation, bengiotranslation}.  The popularity
of DNNs has ushered in the development of several machine learning systems focused on
DNNs~\cite{tensorflow,mxnet,pytorch,caffe2}.  These systems allow users to program a DNN model
with ease in an array-language frontend, and each also enables training on a GPU for
performance.

The power of DNNs lies in their ability to use very large models and to train
with huge datasets.  Unfortunately, such power does not come for free, as
training such networks is extremely time consuming. For example, a single AlexNet
training run takes more than a week using one
GPU~\cite{krizhevsky2012imagenet}.  Compounding the problem, DNN users must
completely re-train the model after any changes in the training parameters while
fine-tuning a DNN.  Given this
frustratingly long training process, the holy grail of a DNN system is to
deliver good training performance across many devices.

The most widely used method for scaling DNN training today is \textbf{data parallelism}.
Traditional DNN training is based on batched stochastic gradient descent where
the batch size is kept deliberately small. Within a batch, computation on each
sample can be carried out independently and aggregated at the end of the batch.
Data parallelism divides a batch among several GPU devices and incurs cross-device
communication to aggregate and synchronize model parameters at the end of each
batch using a parameter service~\cite{muli:osdi14,dean:nn}.  DNN models are large and
growing. For example, in 2012 AlexNet had \mytilde150MB of model parameters, and a
DNN acoustic model from three years later~\cite{dnnacoustic:2015} had 1.6GB of
parameters.  In order to reduce the overhead of synchronizing
large models, one must use very large batch sizes, ensuring that computation
time dominates over communication time.  Data parallelism's reliance on
very large batch size comes at a price: it is known that training using larger
batches converges more slowly, hurts accuracy~\cite{krizhevsky:wierd}, and is more likely to lead
to globally-suboptimal local minima~\cite{largebatch}.

To avoid this \emph{batch-size-dilemma} of data parallelism, one can
divide the model parameters across devices and synchronize the intermediate
computation results instead of the model parameters.  This scheme is referred
to as \textbf{model parallelism}. However, the relative merits of model and data
parallelism depend on the batch size, the model size, and the
shape of the model in use. Previous work~\cite{krizhevsky:wierd} also suggests that
different layers of a DNN model should be treated differently to achieve better training speed.
Due to this unclear trade-off and the fact that
model parallelism requires a more complex implementation, with the exception of
an earlier learning system~\cite{dean:nn}, all existing systems focus on data
parallelism.

Our insight is that data, model and hybrid parallelism can be unified
as approaches to parallelizing tensor operations by
dividing the tensor inputs along different dimensions\footnote{We use the term tensor to refer to multi-dimensional arrays}.
This insight allows 
us to cast the challenge of scaling DNN as a problem of ``tensor tiling'':
\textit{along which dimension should one partition each input or intermediate
	tensor involved in the DNN computation?} As the scaling bottleneck of DNN
training is the communication overhead, we define the corresponding optimization
problem on the tensor dataflow graph, and propose an algorithm to find the
best tiling strategy that minimizes the communication cost. The resulting
tiling strategy is comprehensive. Not only can it partition along 
any one dimension (thus supporting either data or model parallelism), it
also can partition any tensor along more than one dimension (thus supporting a
hybrid version of both data and model parallelism). Furthermore, unlike
existing approaches that adopt a single partitioning strategy for all tensors
involved in the computation, decisions are made separately for each tensor,
depending on its size, shape, and the overall DNN model configuration.

Because of the flexibility allowed for tiling, it is difficult to find an
optimal strategy. In fact, at its full generality, the tiling problem is known
to be NP-complete~\cite{spartan}. Fortunately, many DNN models have the common
structure of multiple stacked neuron layers.  As a result, we can reorganize
the dataflow graph of a DNN training into a chain of levels such that
each level only interacts with the adjacent levels. With this formulation, 
we solve the tiling problem using a novel algorithm that recursively
applies dynamic programming to find the optimal tiling solution given 
any DNN configuration and batch size.

To demonstrate the effectiveness our formulation, we have implemented a
prototype system called \name as a C++-based backend plugin for an existing
DNN system \mxnet. The practice can be applied to any dataflow-based DNN
systems. We evaluated \name on a 8-GPU
machine and compared its performance against data parallelism  with varying
configurations. Our evaluations show that \name is $1.5\times$ to $4\times$ faster
than pure data parallelism for AlexNet and VGG.

\section{Background \& Challenges}
\label{sec:background}
Before introducing \name's approach to efficient deep learning, we must motivate the importance of the problem
and the current popular approaches. We first introduce basic DNN concepts, and describe data parallelism
and model parallelism. We compare the communication costs of these two approaches with a concrete Multi-Layer Perceptron example,
which reveals our core precept: data parallelism, model parallelism, and even hybrid parallelism are just
different ways of tiling tensors in the dataflow graph.
Finally, we discuss the challenges and our contributions of solving this tiling problem in dataflow graph.

\subsection{Background on DNN and Deep Learning Systems}
At its heart, deep learning is an optimization problem, like other machine learning algorithms. Training a DNN 
consists of computing a cost function $C$ given a set of inputs (\emph{forward propagation}), computing the gradient
of that cost function with respect to the model parameters $\theta$ (\emph{backward propagation}), and updating the
model parameters using the gradient just calculated. This training method is called \emph{Stochastic Gradient Descent}
(SGD).  In practice, training is performed by iterating over the training data many times (i.e. ``epoches'') 
doing SGD in mini-batches, as illustrated in the following pseudocode.

\begin{figure}
	\centering
	\includegraphics[width=0.45\textwidth]{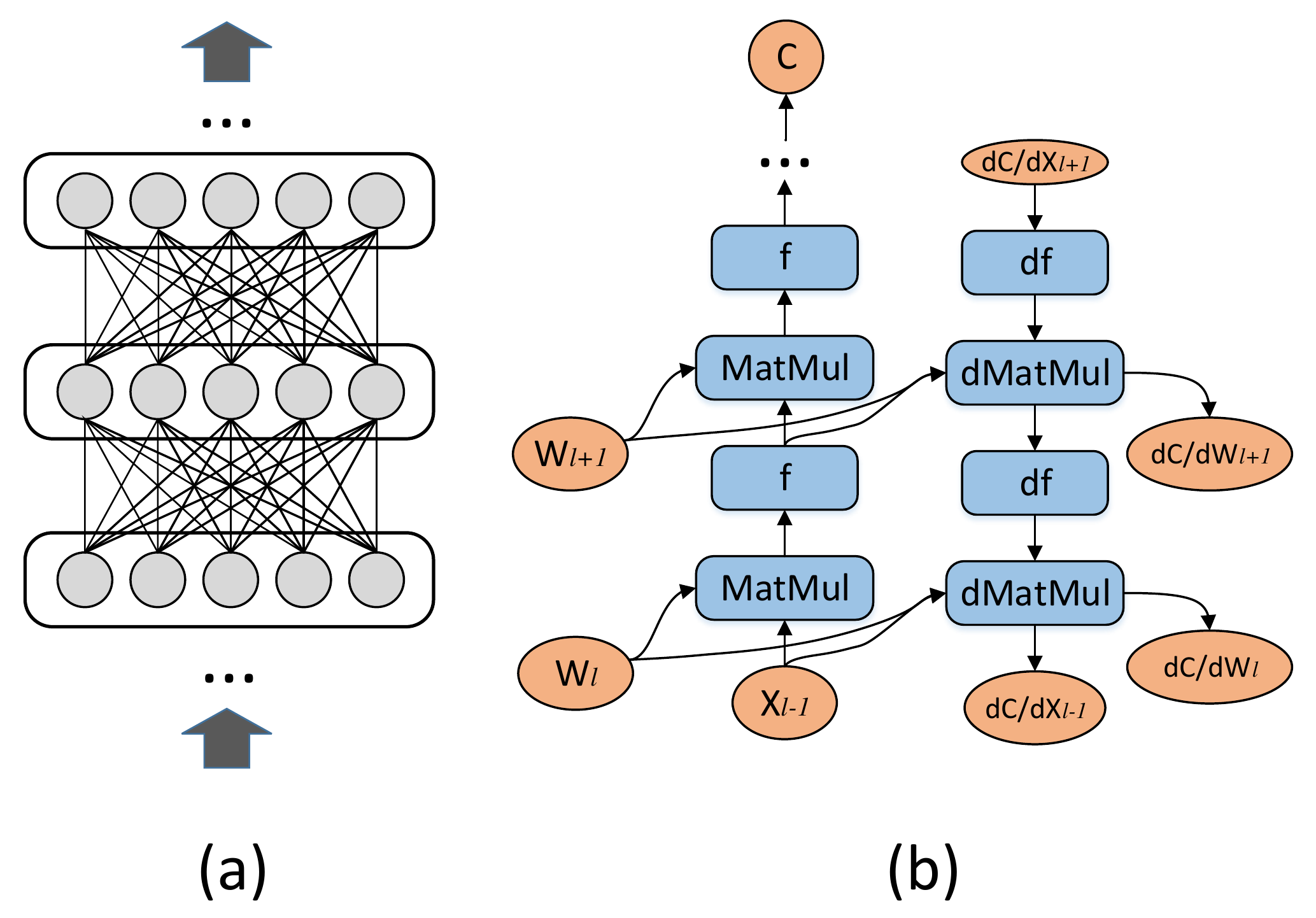}
	\caption{(a) A Multi-Layer Perceptron (MLP) model; (b) The dataflow graph of its forward and backward propagations.}
	\label{fig:mlp}
\end{figure}

\lstdefinelanguage{pseudo}
{morekeywords={for,foreach,to,in,CalculateGradient},
	sensitive=false,
	morecomment=[l]{//},
	morecomment=[s]{/*}{*/}}
\begin{lstlisting}[language=pseudo, mathescape, fontadjust=true, columns=flexible]
for(epoch = 1 to N):
foreach ($D_i$ in $\{D_1, D_2, ...\}$):
$\frac{\partial C}{\partial \theta}$ = CalculateGradient($D_i$, $\theta$)
$\theta := \theta - \epsilon \frac{\partial C}{\partial \theta}$  // update model parameters
\end{lstlisting}

Here, the training dataset $D$ is partitioned into $batches$ $\{D_1, D_2, \ldots\}$ and SGD is performed iteratively,
updating model parameters ($\theta$) after each batch.

DNN models have a common graphical representation, in which layers of neurons
are connected by weighted edges across layers. The weights are the model parameters
to be learned. DNN training involves a series of tensor computations along the
graph structure. Figure~\ref{fig:mlp}(a) shows a Multi-Layer
Perception (MLP) model with 3 fully connected layers.  The weights of neuron connections between successive
layers $l$ and $l+1$ are represented by matrix $W_l$, where $W_l^{i,j}$ is
the weight between the i-th neuron of layer $l$ and j-th neuron of layer $l+1$.
The set of weights of all layers ${W_1, W_2, ..., W_l}$ is referred to as a DNN's model
parameters.

To calculate the cost function, one performs {\em forward propagation} to
compute the \emph{activation} of a layer from its preceding layer.
Specifically, $x_{l+1}=f(x_{l}\cdot W_l)$, where layer $l$'s activation vector
$x_{l}$ is multiplied with the weight matrix $W_{l}$ and then scaled using an
element-wise non-linear function $f$.  The cost function is $C=g(x_{l_{out})}$
where $x_{l_{out}}$ is the activation of the last (outout) layer and $g$ is the loss function.
Gradients are computed through {\em backward propagation} using the chain rule.  The computation proceeds from a higher to
lower layer.  Specifically, two kinds of matrix computations are involved in
each step. One computes the gradient of the activation
$\frac{dC}{dx_{l}}=df(\frac{dC}{dx_{l+1}})W_{l}^T$.  The other computes the
gradient of weight matrix $\frac{dC}{dW_l}=x_{l}^Tdf(\frac{dC}{dx_{l+1}})$.
Here, $df$ is the derivative function of $f$. The weight matrix $W_{l}$ is then
updated using the computed gradient. Our discussion so far performs SGD on one sample only. 
For batched SGD on $b$ samples, the activation of each layer is represented a
matrix of $b$ activation vectors.

Existing deep learning systems such as \tf and \mxnet offer an
array-based frontend for users to express the tensor computation for the
forward propagation.  Because SGD is such a common computation, these systems
automatically derive the computation required for the backward propagation and
handle parameter updates.  The overall computation for both forward and
backward propagation is transformed into a dataflow graph of tensor operators.
As shown in Figure~\ref{fig:mlp}(b),
the dataflow graph for DNN training is mostly serial. Popular DNN systems such as \tf and \mxnet 
directly execute this dataflow on their backends. Hence, if a user wishes
to parallelize training across devices, he or she must manually express the 
parallel computation in the user code so that the resulting dataflow graph
generated has the required parallelism. Furthermore, users are also expected
to explicitly specify placement, i.e. which device should be responsible for
executing which portions of the dataflow graph. This is a tedious process.

\subsection{Scaling challenges and trade-offs}
\begin{figure}[t]
	\centering
	\includegraphics[width=0.5\textwidth]{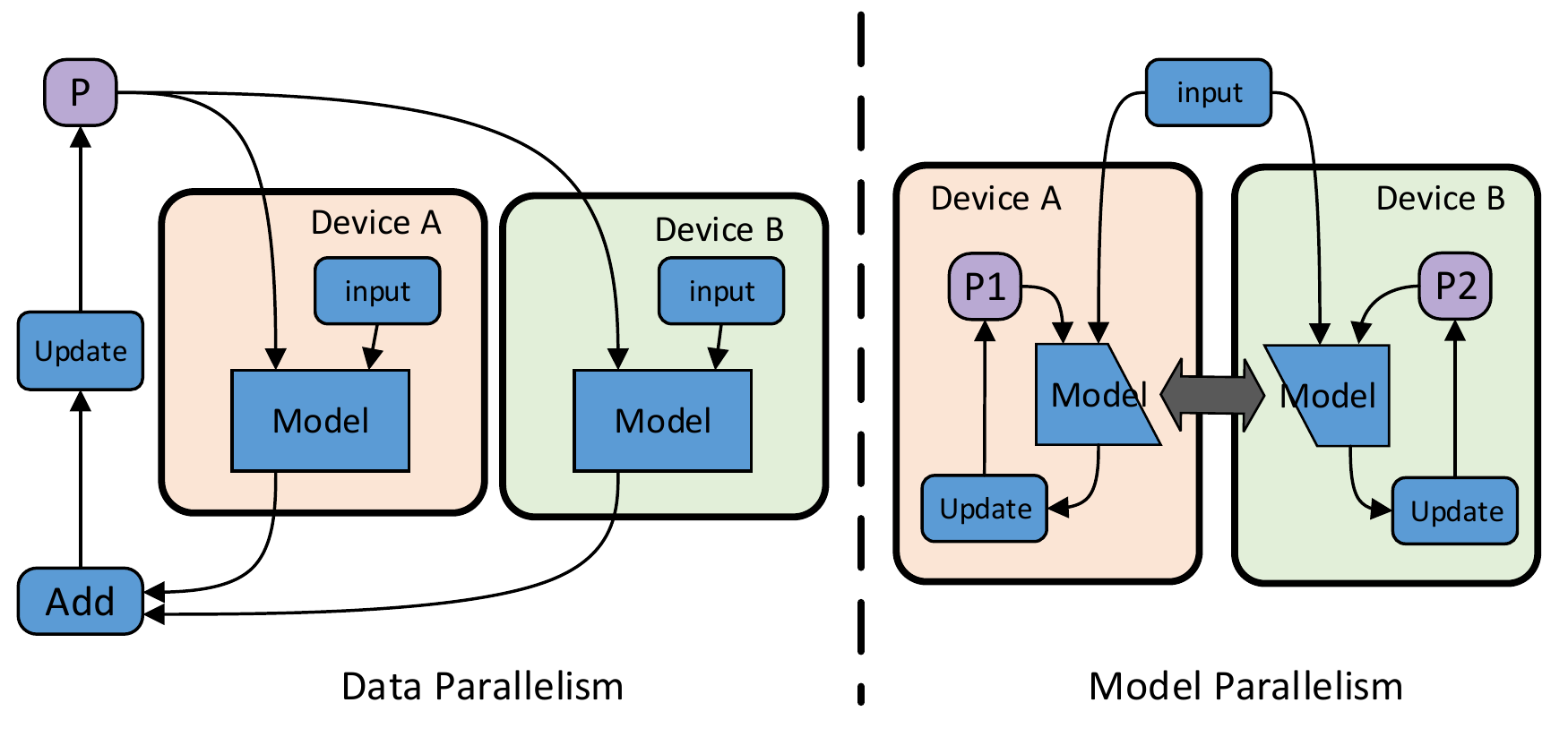}
	\caption{Data parallelism and model parallelism}
	\label{fig:parallelism}
\end{figure}
\textbf{Data parallelism} is the widely-used method for scaling DNN training
across many devices. It takes advantage of the fact that all training samples
in one batch independently contribute to the gradients of the model parameters.
Therefore, data parallelism partitions a batch of samples and lets each device
compute the gradients of the \emph{same} model parameters using a different
partition. The resulting gradients are aggregated before updating the
parameters. The aggregation and update can be done on each device by slicing
parameters evenly, or on a separate device called a \emph{Parameter
	Server}~\cite{muli:osdi14, dean:nn}.  After the model parameters are updated,
they will be \emph{replicated} to all devices for the next iteration. This
results in a Bulk Synchronous Parallel (BSP) approach to parallelizing the
training algorithm (Figure.\ref{fig:parallelism}).

Data parallelism has achieved good speedup for some DNN models (e.g.
Inception network \cite{tensorflow}).  However, since the communication
overhead of data parallelism increases as the model grows bigger, one must
train using a very large batch size to amortize the communication cost across
many devices. In fact, for any DNN model, one can always scale the
``training throughput'' by ever increasing the batch size. Unfortunately, large batch
training is known to be problematic such as longer convergence time
or decreased model accuracy~\cite{krizhevsky:wierd, largebatch}.

\textbf{Model parallelism} partitions the model parameters of each layer among
devices, so that the update of parameters can be performed locally (Figure
\ref{fig:parallelism}). Each device can only calculate part of a
layer's activation using its parameter partition, so all devices need to
synchronize their activations and activation gradients for each layer during
both the forward and backward propagations.
Since model parallelism exchanges activations instead of the
model parameters, it works well for models with small activation
size such as DNN models with large fully-connected layers.



The trade-off between data and model parallelism can be illustrated using an
example as follows. Consider a MLP network with five fully-connected layers; each
layer has 300 neurons; the batch size is 400. Therefore, all weight matrices
are of shape $300\times300$ whereas activation matrices are of shape
$400\times300$ (Figure \ref{fig:paramult}). The model parameter size is
$300\times300\times5\times4\text{B}=1.8\text{MB}$, and the total activation
size of forward propagation $400\times300\times5\times4\text{B}=2.4\text{MB}$.
When parallelizing training for this network on 16 GPUs, data parallelism needs
to first collect all parameter gradients and then synchronize the updated
parameters for all devices, so the total communication is
$1.8\text{MB}\times16\times2=57.6\text{MB}$, while model parallelism transfers
activations and their gradients in both forward and backward propagations, so
the total communication is $2.4\text{MB}\times16\times2=76.8\text{MB}$. Data
parallelism is better than model parallelism in this particular example because
the total activation(gradient) size is bigger than the total parameter size
and data parallelism exchanges parameters across devices. If the batch size is
300 while the layer size is 400, model parallelism becomes better.

However, an even better strategy for this example is a hybrid of data
and model parallelism as follows. The 16 GPUs are divided into four groups.
We use data parallelism among groups, while within each group we apply model parallelism.
The calculation of communication cost is then also divided into two parts.
First, for data parallelism among groups, the communication is
$1.8\text{MB}\times4\times2=14.4\text{MB}$. Second, for model parallelism
within each group, the communication is $\frac{2.4}{4}\text{MB}\times
4\times2=4.8\text{MB}$. Note that the activation size is reduced to
$\frac{2.4}{4}\text{MB}$ due to data parallelism partitioning on the batch
dimension. Finally, because there are four groups, the total communication is
$14.4\text{MB}+4\times 4.8\text{MB}=33.6\text{MB}$. This results in
communication savings of 41.7\% and 56.2\% compared with pure data 
and model parallelism, respectively.

\section{Our approach}
\label{sec:approach}
\begin{figure}
	\centering
	\includegraphics[scale=0.4]{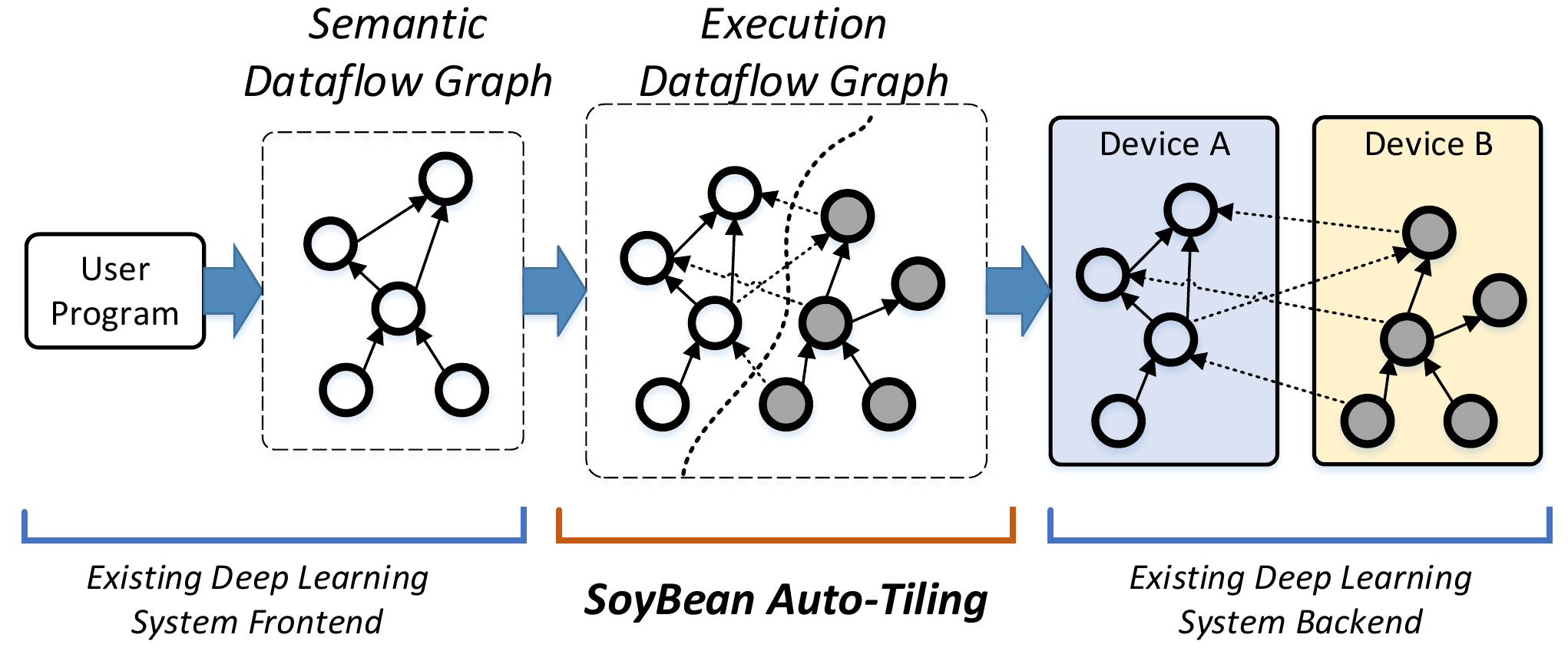}
	\caption{Overview of \name's design.}
	\label{fig:overview}
\end{figure}

Our insight to this challenge is
that data, model and hybrid parallelism can all be unified as
different \emph{tensor tiling schemes} including partitioning a tensor along
certain dimension or replicating the whole tensor. We then find the best parallelism
by finding the tensor tiling scheme that minimizes communication cost.
The formulation lets us support a wide range of parallel strategies used
for DNN training in the literature, including:
\begin{itemize}
	\item Data parallelism. This corresponds to replicating the weight tensors
	and partitioning all other tensors by the data dimension.
	\item Model parallelism. This corresponds to partitioning each weight
	tensor along any one dimension.
	\item Mixed parallelism~\cite{krizhevsky:wierd}. This corresponds to
	distributing some layers using data parallelism and other layers using some model parallelisim strategy. 
	\item Hybrid parallelisim~\cite{dean:nn}.  Here,
	workers are divided into groups. An operator's tensor is first partitioned across groups 
	and then partitioned across workers within a group, using different strategies.
	Combined parallelism has been used by the earlier generation of specialized 
	training systems (e.g. ~\cite{dean:nn, projectadam}), but is not explored
	with the current generation of general-purpose systems due to its programming complexity.
\end{itemize}

We then build a prototype system called \name. Figure.\ref{fig:overview} shows
an overview of \name's overall system design. We reuse the front-end of
existing deep learning systems that express tensor computation by a dataflow graph,
which we refer to as the {\em semantic dataflow graph}. An
example semantic dataflow graph is shown in Figure~\ref{fig:mlp}(b). It is 
mostly serial.  Based on the semantic dataflow graph, \name determines the
best tensor tiling scheme that incurs the least communication cost.
\name then transforms the serial semantic dataflow 
graph into a parallel {\em execution dataflow graph} based on the scheme.
It automatically maps the partitioned arrays and operators to the set of underlying devices.
Finally, the execution graph is dispatched to an existing dataflow engine backend for 
execution. Since the DNN training is done over many iterations using the same
semantic dataflow graph, the runtime cost of the dataflow transformation
can be amortized.

\section{Finding the Optimal Tiling}
\label{sec:algo}

\newcommand{\ttx}{\texttt{X}}
\newcommand{\tty}{\texttt{Y}}
\newcommand{\ttz}{\texttt{Z}}
\newcommand{\tilingx}{t_{\texttt{X}}}
\newcommand{\tilingy}{t_{\texttt{Y}}}
\newcommand{\tilingz}{t_{\texttt{Z}}}
\newcommand{\cost}{c}
\renewcommand{\to}{{\scriptstyle\rightarrow{}}}

Given a dataflow graph, \name aims to find a tiling for \emph{each} tensor in the
graph such that the resulting parallel execution incurs minimal communication
cost. There are three main challenges that \name solves:

\begin{itemize}
	
	\item \emph{Optimize based on the dataflow}: In a dataflow graph, a
	tensor could act as the output of one operator as well as the input of another.
	Hence, a tiling that results in no communication for some operators may lead to
	more communications globally. For example, data parallelism requires no
	communication in forward and backward propagations, but the gradient
	aggregation part may be costly so that the overall communication cost is not
	minimized. \name solves this by considering operators that share inputs or
	outputs together when searching for the optimal tiling. We also show that our
	algorithm could finish in polynomial time thanks to the sequential nature of
	deep learning algorithm.
	
	\item \emph{Determine communication cost}: Given the tilings of the
	inputs and outputs, \name needs to determine the corresponding communication
	costs. In \name, we view communication as tiling conversions. Our core insight
	is that all the data needs to be fetched to the device before the operator can
	be executed. The process of fetching data from one location to another is in
	fact re-organizing the tiles and is thus equal to tiling conversion.
	
	\item \emph{Decompose the optimization problem}: The problem of 
	finding the best tiling for $n$ devices has a very large search space depending on $n$.
	\name provides a recursive solution; it first finds the best tiling for partitioning 
	the tensors among two devices. Then it recursively build upon its baseline
	solution to find tiling for $n>2$ devices.
	
\end{itemize}

In this section, we first discuss the formulation of the tiling problem (Sec~\ref{subsec:tiling}).
Next, we describe how to find the optimal tiling across two devices (Sec~\ref{subsec:one-cut}) and 
build upon this baseline solution to tile across are more than two devices (Sec~\ref{subsec:k-cut}).
We prove the solution's optimality in Sec~\ref{subsec:proof} and discuss extensions in Sec~\ref{subsec:ext}.

\subsection{Parallelism as the Tiling Problem}\label{subsec:tiling}

\begin{figure}[h]
	\centering
	\includegraphics[scale=0.53]{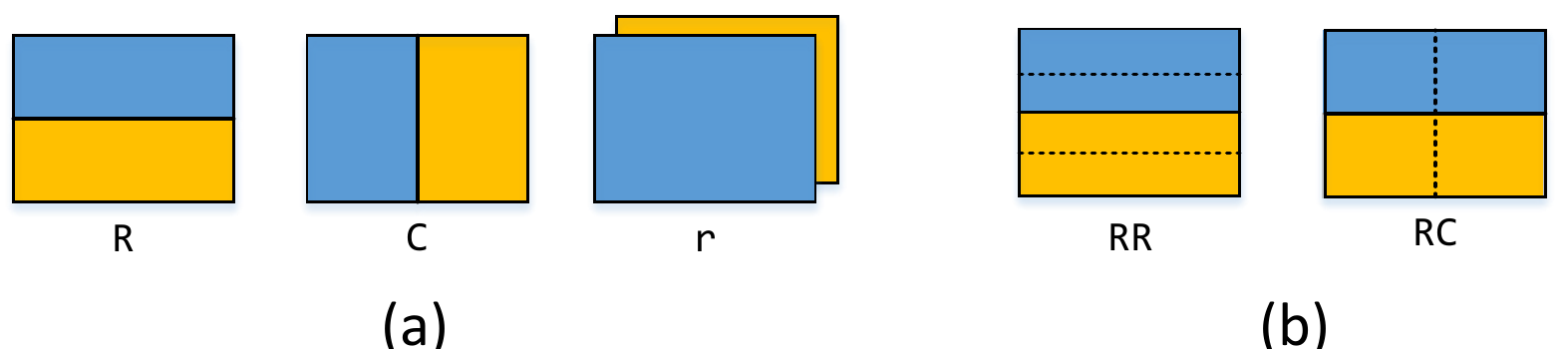}
	\caption{(a) Three basic tilings: row tiling, column tiling and replication; (b) Left: compose two basic row tilings into a four-way row tiling; right: compose row tiling and column tiling to partition matrix into four blocks.}
	\label{fig:basic-schemes}
\end{figure}

We formulate the tiling problem to be solved by \name.  To ease the discussion,
we only consider 2-D tensor (matrix) here. Extension to
high-dimensional tensor is straight-forward and will be covered in section
\ref{subsec:ext}. 

We first define the set of supported tiling schemes of a matrix.  \name only
considers \emph{even} tiling schemes that result in (sub-)tensors of the same shape
and size, because we want to balance the computation across all devices.  
There are three basic tilings that divide a matrix computation into two equal parts:
\textit{row tiling}, \textit{column tiling} and \textit{replication}, as shown in
Figure.\ref{fig:basic-schemes}(a). Since all tiles are of the same shape, basic
tillings can be applied again on each tile to further partition the matrix.
We call this \emph{tiling composition}. The result of a tiling composition is
still an even tiling.

Let $T^1=\{\schR, \schC, \schr\}$ be the set that contains all basic tilings of
a matrix, where \schR, \schC{} and \schr{} represent row tiling, column tiling
and replication, respectively. We then define a $k$-cut tiling set that
contains all possible tilings after $k$ compositions as follows:

\begin{defi}
	$T^k=\{ t_1t_2| t_1\in T^{k-1}\wedge t_2\in T^1 \},~\forall k\geqslant2$.
\end{defi}
For example, a 2-cut tiling set $T^2$ is as follows:
\[
T^2=\{\schR\schR, \schR\schC, \schR\schr, \schC\schR, \schC\schC, \schC\schr, \schr\schR, \schr\schC, \schr\schr \}
\]
Figure.\ref{fig:basic-schemes}(b) shows how 2-cut tilings \schR\schR{} and
\schR\schC{} partition the matrix into four pieces. Note that a $k$-cut tiling
partitions a matrix into $2^k$ pieces.  To simplify the discussion, we assume
the number of workers is $n = 2^k$.

We then formally define the tiling of a dataflow graph:
\begin{defi}
	The $k$-cuts tiling of a dataflow graph $G$ is a function $\mathcal{T}_G: M\mapsto T^k$ that maps from all the matrices $M$ in dataflow graph to their tilings.
\end{defi}

\begin{figure}
	\centering
	\includegraphics[scale=0.6]{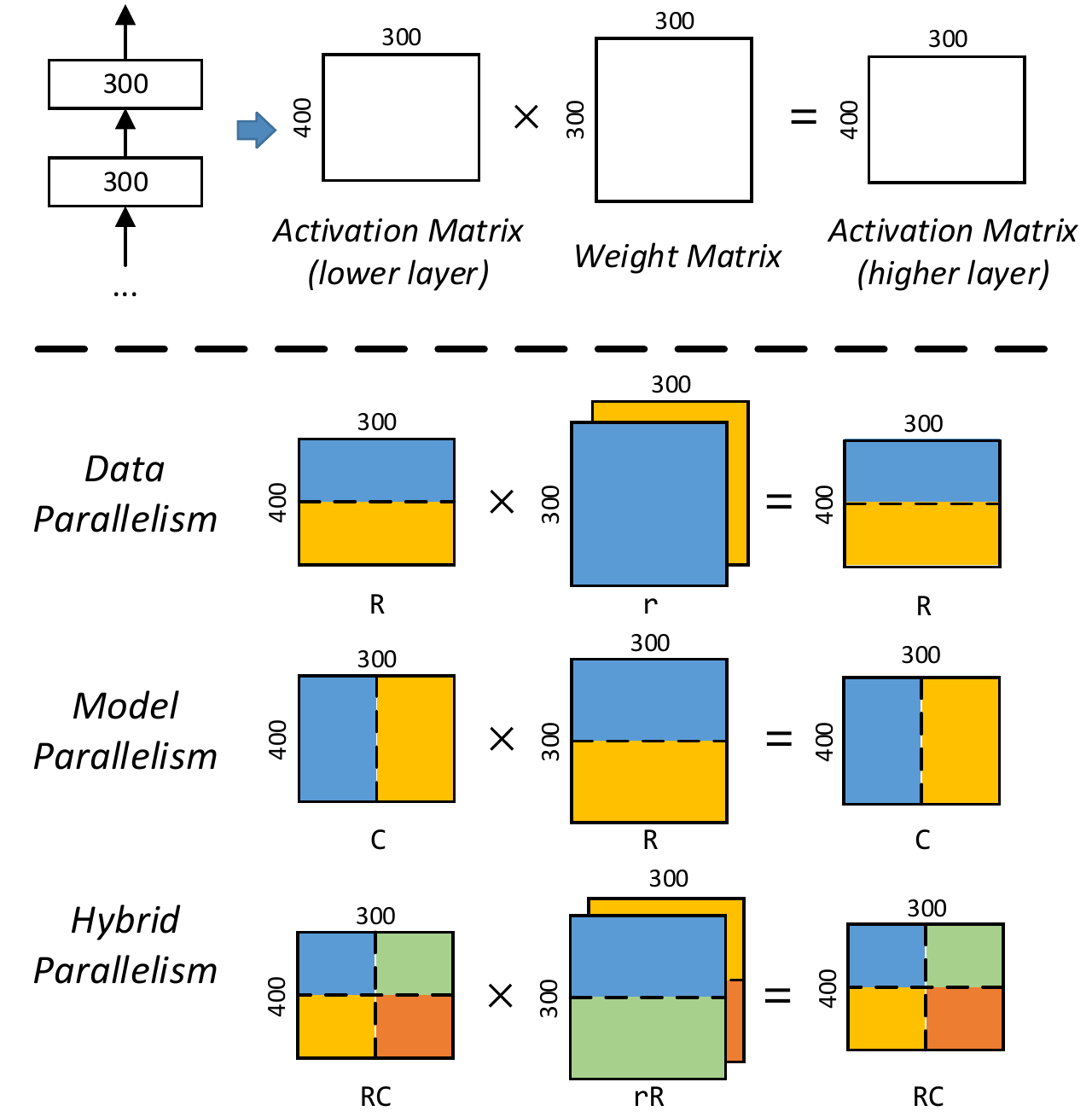}
	\caption{Top: forward propagation of one layer in a MLP model. Bottom: how matrices are tiled in the forward propagation for different parallelisms.}
	\label{fig:paramult}
\end{figure}

Given the above definition, we could unify data, model and hybrid parallelisms
as different tilings of the dataflow graph (Figure.\ref{fig:paramult}).  We
explain why this is the case using the same MLP example in
Figure.\ref{fig:mlp}. 
\begin{itemize}
	\item In data parallelism, all activations are partitioned along the batch dimension while all parameters are replicated. Suppose the batch dimension is the row dimension. Then data parallelism on two devices can be described by the following tiling:
	\[
	\mathcal{T}_{data}(m)=
	\left\{
	\begin{array}{ll}
	\schr&\text{if }m\text{ is weight matrix},\\
	\schR&\text{otherwise.}
	\end{array}
	\right.
	\]
	\item In model parallelism, the weight matrices are partitioned in order to avoid gradient aggregation among devices. For a 2D weight matrix, there are two possible tilings: \schR{} and \schC{}. Suppose we choose row tiling. In order to compute the gradient of parameters locally, the activation matrices are partitioned along column while the activation gradients are replicated. Therefore, model parallelism on two devices can be described as following tiling:
	\[
	\mathcal{T}_{model}(m)=
	\left\{
	\begin{array}{ll}
	\schR&\text{if }m\text{ is weight matrix},\\
	\schC&\text{if }m\text{ is activation},\\
	\schr&\text{otherwise.}
	\end{array}
	\right.
	\]
	\item Hybrid parallelism divides all devices into groups. It supports one type of parallelism within a group and a different type across groups. This can be described as a \emph{composition} of $\mathcal{T}_{data}$ and $\mathcal{T}_{model}$. An example tiling of hybrid parallelism on four devices is as follows. It performs data parallelism across groups and model parallelism within a group.
	\[
	\mathcal{T}_{hybrid}(m)=
	\left\{
	\begin{array}{ll}
	\schr\schR&\text{if }m\text{ is weight matrix},\\
	\schR\schC&\text{if }m\text{ is activation},\\
	\schR\schr&\text{otherwise.}
	\end{array}
	\right.
	\]
\end{itemize}

\subsection{Tiling across two devices}\label{subsec:one-cut}

We first solve the base case problem of finding the best tiling for two devices
that minimizes communication. Let us consider the
MLP model in Figure~\ref{fig:mlp}(b). If there are $L$ layers, the forward
propagation for computing loss value $\mathcal{L}$ can be simplified as follows:
\begin{equation}
\mathcal{L}=x_0\prod_{1\leqslant l\leqslant L}W_l
\end{equation}
Here, we ignore the element-wise functions. $x_0$ is the input data and $W_l$
is the weight matrix of layer $l$. \emph{How to assign the tiling schemes
	for all matrices in this computation to minimize the communication cost?}
We first explain how to calculate the communication cost of one
matrix multiplication given the tilings of its input/output matrices (Sec~\ref{sec:cost}).
Then we give an algorithm to optimize for the cost (Sec~\ref{sec:one-cut-algo}).

\subsubsection{Calculating the communication cost}\label{sec:cost}
\begin{figure}
	\centering
	\includegraphics[scale=0.4]{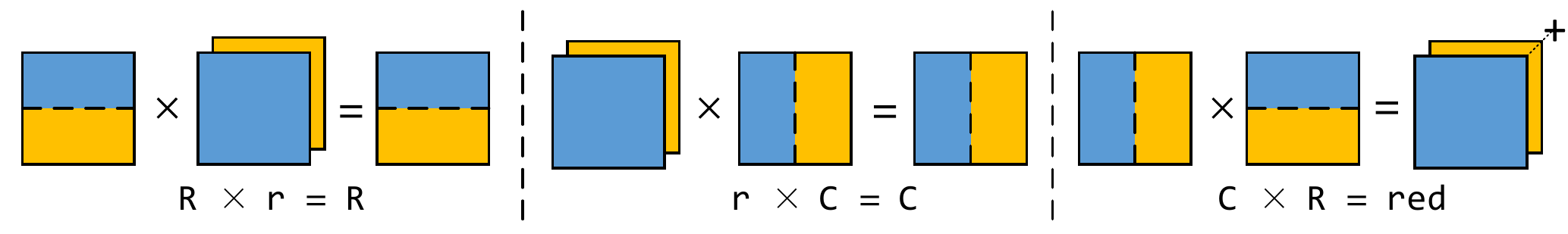}
	\caption{Three forms of aligned tilings for matrix multiplication.  The resulting partition of the third form is an intermediate one $\schred$, and requires an extra reduction.}
	\label{fig:matmult}
\end{figure}
\begin{figure}[t]
	\centering
	\includegraphics[width=0.4\textwidth]{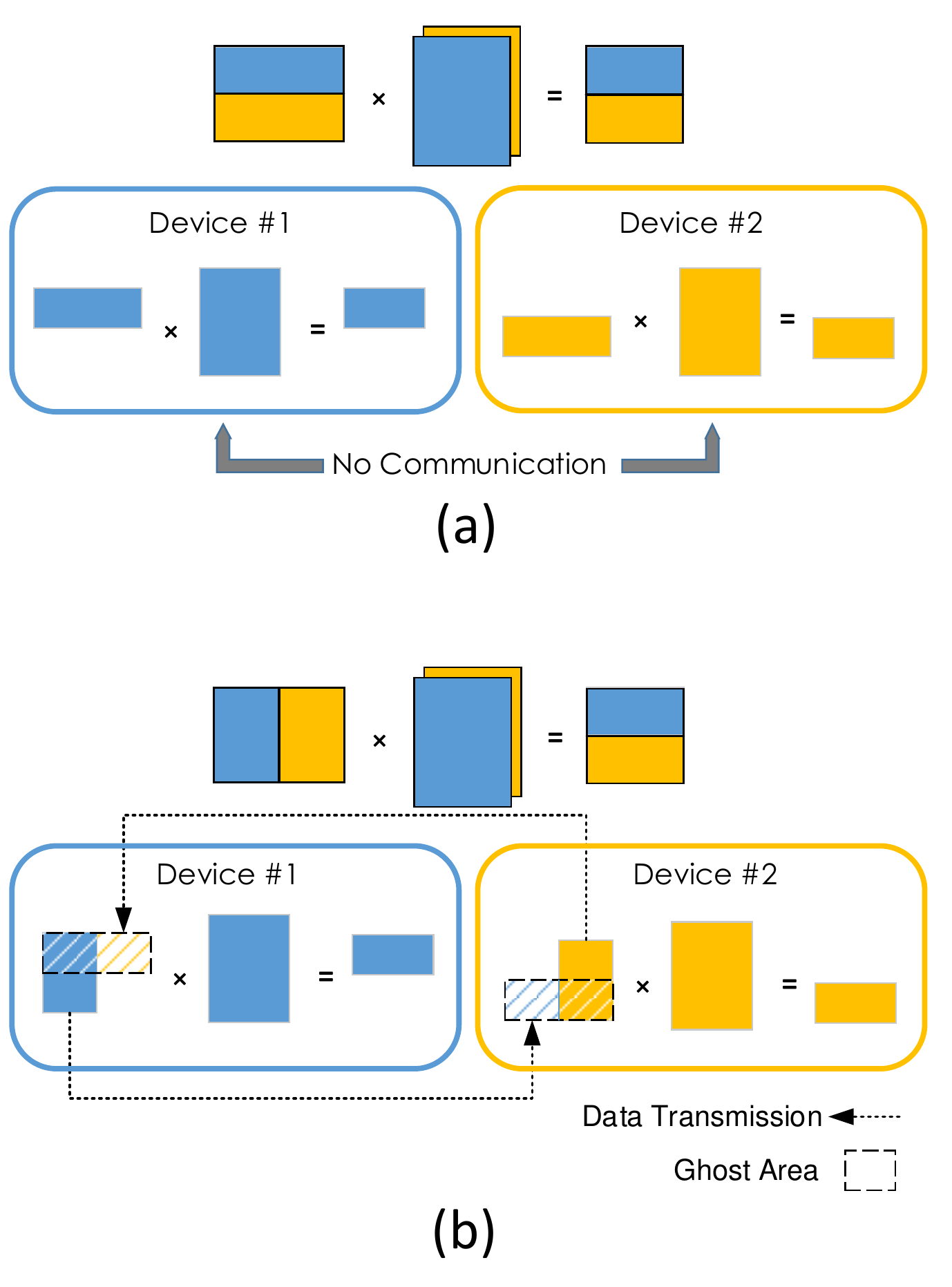}
	\caption{(a) \emph{Aligned} multiplication that has no communication. (b) \emph{Unaligned} multiplication. The dashed areas are ``ghost areas" representing the area after tiling conversion. The yellow area on device \#1 needs be fetched from device \#2.}
	\label{fig:align}
\end{figure}
Let us first consider only one matrix multiplication, $\ttx\times \tty=\ttz$. What is the communication cost given the tilings of its inputs $\tilingx$ and $\tilingy$ and its output $\tilingz$?

This problem is a variant of the traditional block matrix multiplication, except in addition to being partitioned along rows and columns, the matrix can also be replicated. The core principle here is that each submatrix multiplication cannot be performed without all its inputs being fetched to the device on which it is executed. Therefore, we can view a parallel matrix multiplication as having three phases:
\begin{itemize}
	\item \emph{Inputs Conversion Phase}: The tiles of the inputs ($\ttx$ and $\tty$) are fetched to the device(s) that require them for computation, if they reside on different devices.
	\item \emph{Computation Phase}: Submatrix multiplications are executed on all devices locally.
	\item \emph{Outputs Conversion Phase}: The temporary outputs of the local matrix multiplications are pushed to the devices specified by the tilings of the outputs ($\ttz$).
\end{itemize}

Notably, communication only happens in the Inputs and Outputs Conversion
phases, which means we can compute the communication cost by computing the
conversion costs of different tilings.  However, since there are many types of
input and output tilings, we do not want to enumerate all possible combinations
to calculate the cost.  Rather, we find a small set of so-called {\em aligned
	tiling} and reduce all tiling combinations to one of those cases.  Aligned 
tilings may or may not correspond to real tilings.  We use them to simplify the
conversion cost calculation by drastically cutting down the number of cases
we have to consider.

Given the three basic tilings, there are three corresponding aligned
tilings, as depicted in Figure.\ref{fig:matmult}.  For the left most
tiling, the two inputs are row-partitioned and replicated, respectively and  
output is row partitioned.  Similarly for the scenario in the middle. 
There is no communication cost for these two tilings.  The third scenario is
different in that its output tiling does not correspond to any basic tiling.  Rather, the intermediate matrices computed on 
each device need to be aggregated later using an extra reduction operation.
We denote this intermediate tiling as $\schred$.

All of the aligned tilings
have several properties in common. First of all, they are all \emph{correct}
block matrix multiplications. Second, removing any submatrix product will give
wrong results. Therefore, there is no redundant computation. Finally, they are
\emph{balanced} in that the number of submatrix products is equal to the number
of devices. 

\paragraph{One-cut Communication Cost:}
Let us define $\cost(t_1\to t_2)$ as the conversion cost from tiling $t_1$ to $t_2$. To compute the communication cost of a matrix multiplication, we calculate the minimum cost of converting the given input/output tilings to one of the three aligned tilings in Figure~\ref{fig:matmult}.  To put it formally, the communication cost $\cost(\tilingx, \tilingy, \tilingz)$ of a matrix with input tiling $\tilingx$, $\tilingy$ and output tiling $\tilingz$ is as follows:
\begin{equation}\label{eq:multcost}
\min\left\{
\begin{array}{l}
\cost(\tilingx\to \schR) + \cost(\tilingy\to \schr) + \cost(\schR \to \tilingz),\\
\cost(\tilingx\to \schr) + \cost(\tilingy\to \schC) + \cost(\schC \to \tilingz),\\
\cost(\tilingx\to \schC) + \cost(\tilingy\to \schR) + \cost(\texttt{red} \to \tilingz)
\end{array}
\right\}
\end{equation}
Computing the cost of tiling conversion is straightforward. It is equal to the area required for the local multiplication (which we refer to as \emph{``ghost area"}) minus the area that already exists on the device. Figure~\ref{fig:align}(b) illustrates how an unaligned multiplication $\schC\times\schr=\schR$ is computed through a conversion to an aligned multiplication $\schR\times\schr=\schR$. The submatrix required for local computation  is marked by the dashed line. The submatrix filled by yellow shading on device \#1 needs to be fetched from device \#2. Hence, its area is equal to the amount of communications involved in this tiling conversion.

\subsubsection{One-cut tiling algorithm}\label{sec:one-cut-algo}
Given a dataflow graph $G$, the one-cut tiling algorithm finds a tiling across
two devices (or groups), $\mathcal{T}_{min}:M\mapsto T^1$, such that the
overall communication cost is minimized:

\begin{equation}
\mathcal{T}_{min}=\argmin_{\mathcal{T}}\sum_{o\in O_G}\cost(\mathcal{T}(o_{\ttx}), \mathcal{T}(o_{\tty}), \mathcal{T}(o_{\ttz}))
\end{equation}
where $O_G$ represents all the matrix multiplications in the dataflow graph $G$, and $o_{\ttx}$, $o_{\tty}$ and $o_{\ttz}$ represent the input matrices and output matrix of a matrix multiplication $o$.

The central problem in searching optimal tilings is that changing the tiling of one matrix may affect many multiplications. Again, let us look at the MLP example. If we only consider the forward propagation part $\mathcal{L}=x_0\prod_{1\leqslant l\leqslant L}W_l$, we can first calculate the communication cost of $x_1=x_0W_1$ under all the possible tilings of $x_0$, $x_1$ and $W_1$. We then proceed to the next multiplication $x_2=x_1W_2$, but when choosing $x_1$ to be tiling $t_1$, we should include the minimal cost of $x_1$ to be $t_1$ in the first multiplication. The total communication cost is then the minimal cost after we finish the last multiplication. When further taking the backward propagation part $\frac{d\mathcal{L}}{dx_0}=\frac{d\mathcal{L}}{dx_L}\prod_{L\geqslant l\geqslant 1}W_l^T$ into consideration, two multiplications ($x_l=x_{l-1}W_l$ and $\frac{d\mathcal{L}}{dx_{l-1}}=\frac{d\mathcal{L}}{dx_l}W_l^T$) should be considered together, because the tiling of $W_l$ affects both multiplications.

Our tiling algorithm exploits an intuitive idea: operators that share inputs or outputs should be considered together. To achieve this, our algorithm first treats the dataflow graph as an undirected graph $G'$, and then uses a breadth-first search on this graph to organize graph nodes into a list of levels $L=\langle l_0,l_1,\ldots,l_n\rangle$. BFS puts nodes that share inputs or outputs in adjacent levels. We then use dynamic programming (DP) to search for optimal tilings:

\noindent{}\textit{Initial condition:}
\begin{equation}
g_0(\tau_0)=\text{level\_cost}_0(\phi, \tau_0)
\end{equation}
\noindent{}\textit{DP equation} ($l\geqslant 1$):
\begin{equation}
g_l(\tau_l)=\min_{\tau_{l-1}}\left\{ \text{level\_cost}_l(\tau_{l-1}, \tau_l) + g_{l-1}(\tau_{l-1}) \right\}
\end{equation}
Here, $\tau_l$ contains the tilings of all matrices that are shared among multiplications in level $l$ and $l+1$. The $\text{level\_cost}_l$ is calculated by summing up the cost of each matrix multiplication in level $l$. The DP state $g_l(\tau_l)$ represents the minimal communication cost after executing all the operators from level zero to level $l$ and also partitioning matrices used by level $l+1$ by tilings in $\tau_l$.

The algorithm searches for all possible tilings and therefore guarantees optimal. Unfortunately,
the running time of the DP algorithm is exponential. Because all the operations are matrix multiplications,
the node degree in the undirected graph $G'$ is at most three. According to Moore bound, if there are $N$ nodes
in the graph, the diameter is $O(\log N)$. Since the number of BFS levels is equal to the graph diameter,
the maximum number of nodes in each level is $O(N/log N)$. Computing the total cost of each level needs to explore
all tiling combinations of the inputs and outputs of the operators in that level. Therefore, the
worst-case complexity of calculating $\text{level\_cost}_l$ is $|T^1|^{O(N/\log N)}=O(3^N)$.

The observation that saves us is that deep learning does not have an arbitrary
dataflow graph.  Rather, its graph usually has a large diameter. This is because training
neural network usually involves computing each layer sequentially. For an
$N$-layer MLP network, there are $3N$ matrix multiplications (forward,
backward and gradient computation) and the diameter is $N$. In this case, the
average number of nodes in each level is some constant $c$, so the average
running complexity of the whole DP algorithm is linear, $O(3^c N)$.

\subsection{Tiling across many devices}\label{subsec:k-cut}
The $k$-cut algorithm finds the optimal tiling for $n=2^k$ devices.
It is a recursive algorithm.  Specifically, we can divide 
$2^k$ devices into 2 groups, each with $2^{k-1}$ devices.
We first use the one-cut algorithm to find the 
best tiling to partition the computation among the two groups.
Within each group, we perform $(k-1)$-cuts to find the optimal tiling.
Algorithm~\ref{alg:kcuts} shows the pseudocode.

\begin{algorithm}[t]
	\SetKwFunction{OneCutTiling}{OneCutTiling}
	\SetKwFunction{KCutsTiling}{KCutsTiling}
	\SetKwFunction{Construct}{ConstructTileGraph}
	\SetKwInOut{Input}{input}\SetKwInOut{Output}{output}
	
	\Input{A dataflow graph $G$ and a number $k$}
	\Output{A $k$-cuts tiling $\mathcal{T}$ of the graph and its communication cost $c$}
	
	\If{$k=0$}{\Return $\phi$ and $0$}
	\Else{
		$\mathcal{P}_k$, $\delta_k$ $\leftarrow$ \OneCutTiling{$G$}\;
		$G'$ $\leftarrow$ \Construct{$G$, $\mathcal{P}_k$}\;
		$\mathcal{T}_{k-1}$, $\cost_{k-1}$ $\leftarrow$ \KCutsTiling{$G'$, $k-1$}\;
		$\mathcal{T}_k$ $\leftarrow$ $\mathcal{P}_k\circ\mathcal{T}_{k-1}$\;
		$\cost_k$ $\leftarrow$ $\delta_k + 2\cost_{k-1}$\;
		\Return $\mathcal{T}_k$ and $c_k$
	}
	\caption{$k$-cuts tiling algorithm}\label{alg:kcuts}
\end{algorithm}

In Algorithm~\ref{alg:kcuts}, $\mathcal{P}_{k}\circ\mathcal{T}_{k-1}$ is the composition of two tilings (see Section \ref{subsec:tiling}). $\delta_k$ is the cost of the $k$-th cut. As each cut partitions computation into two groups, the cost of the subproblem is multiplied by two. The total communication cost is the weighted summation of per-cut costs:
\begin{thm}[Total communication cost]\label{thm:total}
	$c_{k}=\sum_{i=1}^{k}2^{k-i}\delta_i$
\end{thm}

\subsection{Proof of Optimality}\label{subsec:proof}
Due to the limit of space, we only emphasize the key point that leads our proof. The core property that makes the $k$-cuts algorithm optimal is the \emph{commutativity} of tiling composition because they are orthogonal ways of partitioning. If we let $\schR^k$ be the $k$-row tiling, then the $T^2$ tiling set could be rewritten as $T^2=\{\schR^2, \schC^2, \schr^2, \schR\schC, \schR\schr, \schC\schr \}$. In fact, we have:
\begin{thm}[Flattening]\label{thm:flattening}
	\normalfont
	\[
	T^k=\{\schR^{k_\schR}\schC^{k_\schC}\schr^{k_\schr}|k_\schR,k_\schC,k_\schr\in \mathbb{Z}\wedge k_\schR+k_\schC+k_\schr=k \}
	\]
\end{thm}
Let $\mathcal{T}_k=\langle\mathcal{P}_k, \mathcal{P}_{k-1}, \ldots, \mathcal{P}_{1} \rangle$ be the tiling sequence generated by our $k$-cuts algorithm. We can also prove that the order of the tiling applied will not influence the total communication cost $c_k$. This gives us the following property:
\begin{thm}[Greediness]\label{thm:greedy}
	Let $\delta_{k}, \delta_{k-1}, \ldots \delta_1$ be the cost of each tiling. We have $\delta_i\leqslant2\delta_{i-1},~\forall 2\leqslant i\leqslant k$.
\end{thm}
The greedy theorem means that for any tiling sequence $\mathcal{T}_k$ we can reorder it such that the contribution of the cost of each tiling is increasing. Suppose there is another tiling sequence $\mathcal{T}_k'$ that is not chosen by our algorithm but has $c_k'<c_k$, we can prove that there must exist a ``cross" step such that after that step, the remaining tilings in $\mathcal{T}_k$ produces larger communication cost than the remaining tilings in $\mathcal{T}_k'$. We can prove that by choosing the tiling at the ``cross" step in $\mathcal{T}_k'$, it will give smaller cost than the one in $\mathcal{T}_k$ which contradicts to the local optimality of $k$-cuts algorithm.

\subsection{Extensions to General Dataflow Graph}\label{subsec:ext}
First, we extend the $k$-cuts tiling algorithm to high-dimensional tensors.
We change the basic tiling set to $T^1=\{\schP_1, \schP_2, \ldots,
\schP_d, \schr\}$, where $\schP_d$ represents partitioning along $d^{th}$ dimension. The running
complexity of the one-cut algorithm then becomes $O((d+1)^cN)$ given $N$ nodes
in the dataflow graph, which is still acceptable provided that $d$ is usually
small.

Second, we discuss how to handle operators beyond matrix multiplications.
Recall that the communication cost is equal to the tiling conversion cost.
The only information that is tied to operator type is the set of
the aligned tilings of an operator. Here, we categorize operators for discussion. For
element-wise functions (e.g. non-linear activation function), the only aligned
computation method is to have the same tiling for all of the operator's inputs
and outputs. Note that having all of an operator's inputs and outputs
replicated is not allowed due to redundant computation. For 
convolution, the activation and parameter tensors have four dimensions. Tiling
on batch dimension leads to data parallelism while tiling on channel dimensions
leads to model parallelism. Tilings on image and kernel dimensions are strictly
worse than data parallelism so is ignored in our implementation. Note that the
image and kernel sizes are in fact multipliers on the batch size and weight
size, respectively. The larger the image size, the larger the activation tensor
and thus the better data parallelism is. Finally, for all other operators, we
only allow partitioning on the batch dimension, resulting in data parallelism.
These changes should be easy since they only focus on the operator semantics
while the complicated communication patterns will be automatically deduced by \name.

\section{Constructing the Execution Dataflow Graph}
This section covers how \name dispatches operators to different devices and
how the semantic dataflow graph is converted to the execution graph given the $k$-cuts
tiling schemes computed by our algorithm.

\subsection{Tile Placement}
The first consideration is load balancing. Fortunately, \name's tiling algorithm
ensures that all tensors and operators in the dataflow graph are evenly partitioned,
which means the workload is perfectly balanced.

Another consideration is the interconnects between devices. For example, GPUs within
a single machine can be connected to different CPUs by PCI-e. Given the fixed total
amount of communications, we want the majority of data transmission to be between
GPUs attached to the same CPU to avoid the high latency of QPI connections.
When scaling beyond one machine, this becomes more critical since network transmission is even slower.
At a high level, the interconnection hierarchy divides devices into groups and encourages communications within each group.
The $k$-cuts tiling algorithm naturally fits this scenario because each cut
partitions the workload into two groups of devices; each group is then recursively partitioned.
Theorem \ref{thm:greedy} indicates that \name tends to partition groups
such that the majority of communication happens within groups.
Therefore, our placement strategy follows the algorithm structure.
We map the workloads partitioned by the first cut to the groups connected
by the slowest interconnects (e.g. Ethernet or QPI). We then continue mapping
workloads within each group to the second slowest interconnects, and so on.

\subsection{Connecting Partitioned Operators}
The $k$-cuts tiling algorithm partitions each operator in the semantic graph into $2^k$ sub-operators.
The inputs and outputs of these sub-operators are tiles (sub-tensors) of their original inputs
and outputs. Therefore, to construct the execution graph is to connect the sub-operators for
every connected operators in the semantic graph. Similar to the aligned matrix multiplication,
this includes three phases:
\begin{enumerate}
	\item We need to first convert the input tiling to the aligned tiling for
	the downstream operator. The tiling conversion dispatches each tile to the
	location specified by the placement. A simple solution is letting the receiver
	devices pull from the device that contains the area they need. The situation
	becomes complicated when the receiver only needs one slice of the sender's data
	and needs to concatenate slices from multiple senders. In this case, we transfer
	the data in three steps. First, the flattening theorem (Theorem \ref{thm:flattening})
	allows us to represent each original tensor as a grid of tiles. Therefore,
	the sender can slice its own tile into shards such that each shard is dispatched
	to different receiver. Second, the receivers will fetch the shards they need from senders.
	Finally, the flattening theorem again allows the receiver to directly concatenate the
	received shards back to tiles.
	\item Once the tiling conversions of inputs are done, all sub-operators can be
	executed locally by the definition of aligned tiling. Moreover, since all sub-operators
	are identical, we can just connect the input tiles to the sub-operators one by one.
	\item The temporary outputs of the sub-operators again should be converted to the output
	tilings for later computation. This is the same process as the input tiling conversion.
\end{enumerate}

\section{Evaluation}
\begin{figure*}[t]
	\begin{center}
		\subfloat[Batch size = 512; Hidden Size = 8K]{%
			\includegraphics[width=.33\textwidth]{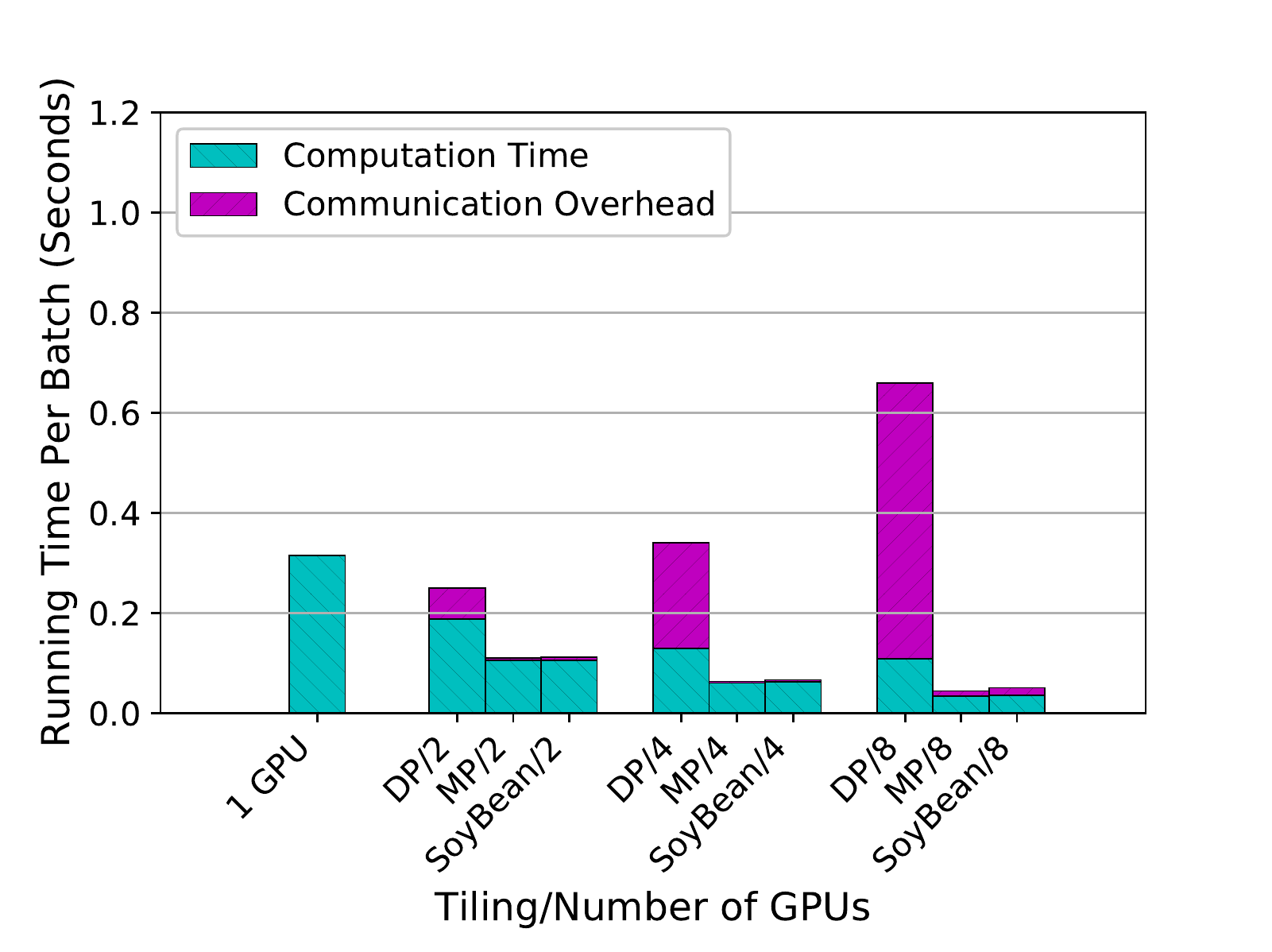}}
		\subfloat[Batch size = 2K; Hidden Size = 8K]{%
			\includegraphics[width=.33\textwidth]{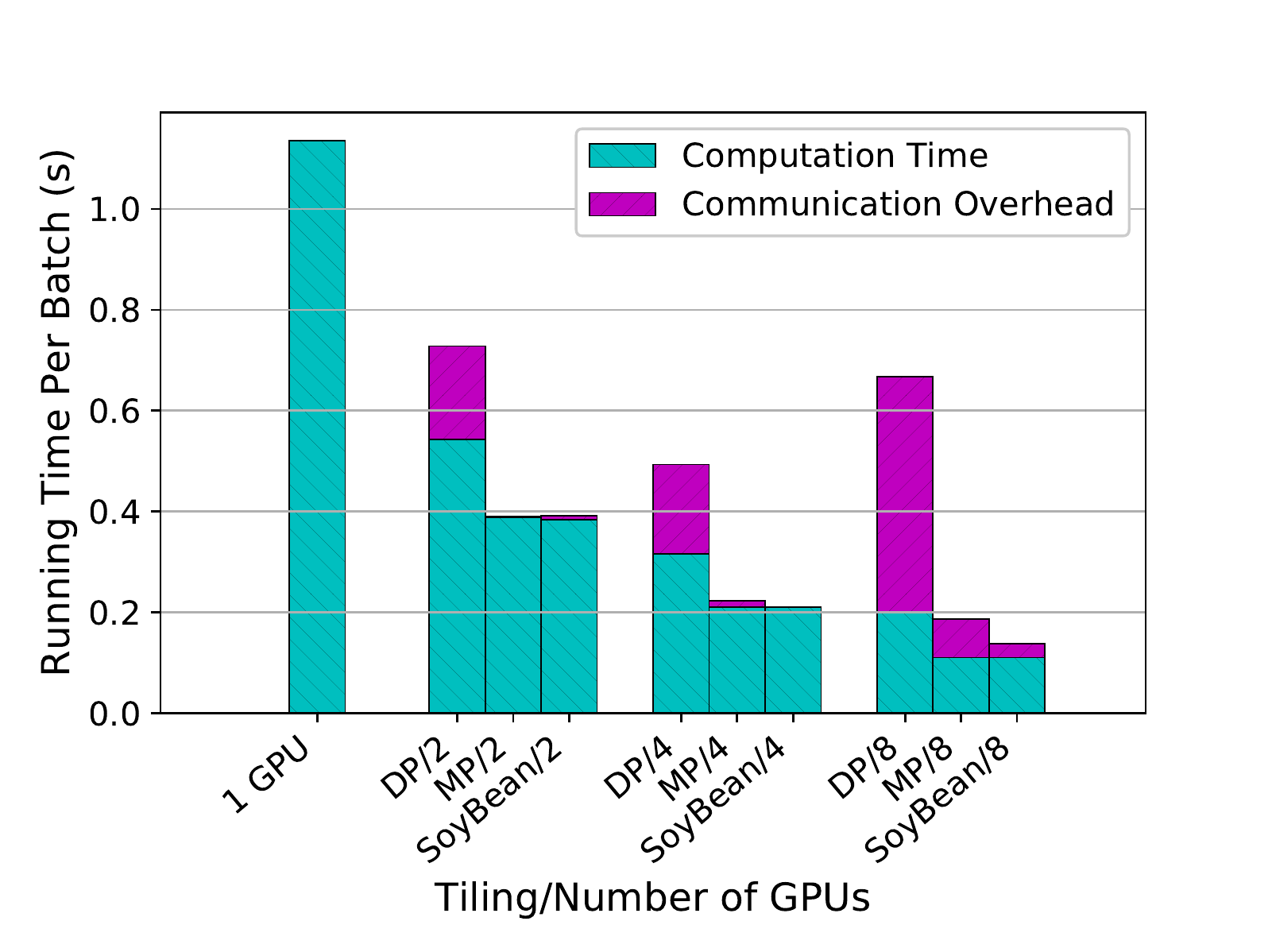}}
		\subfloat[Batch size = 2K; Hidden Size = 12K]{
			\includegraphics[width=.33\textwidth]{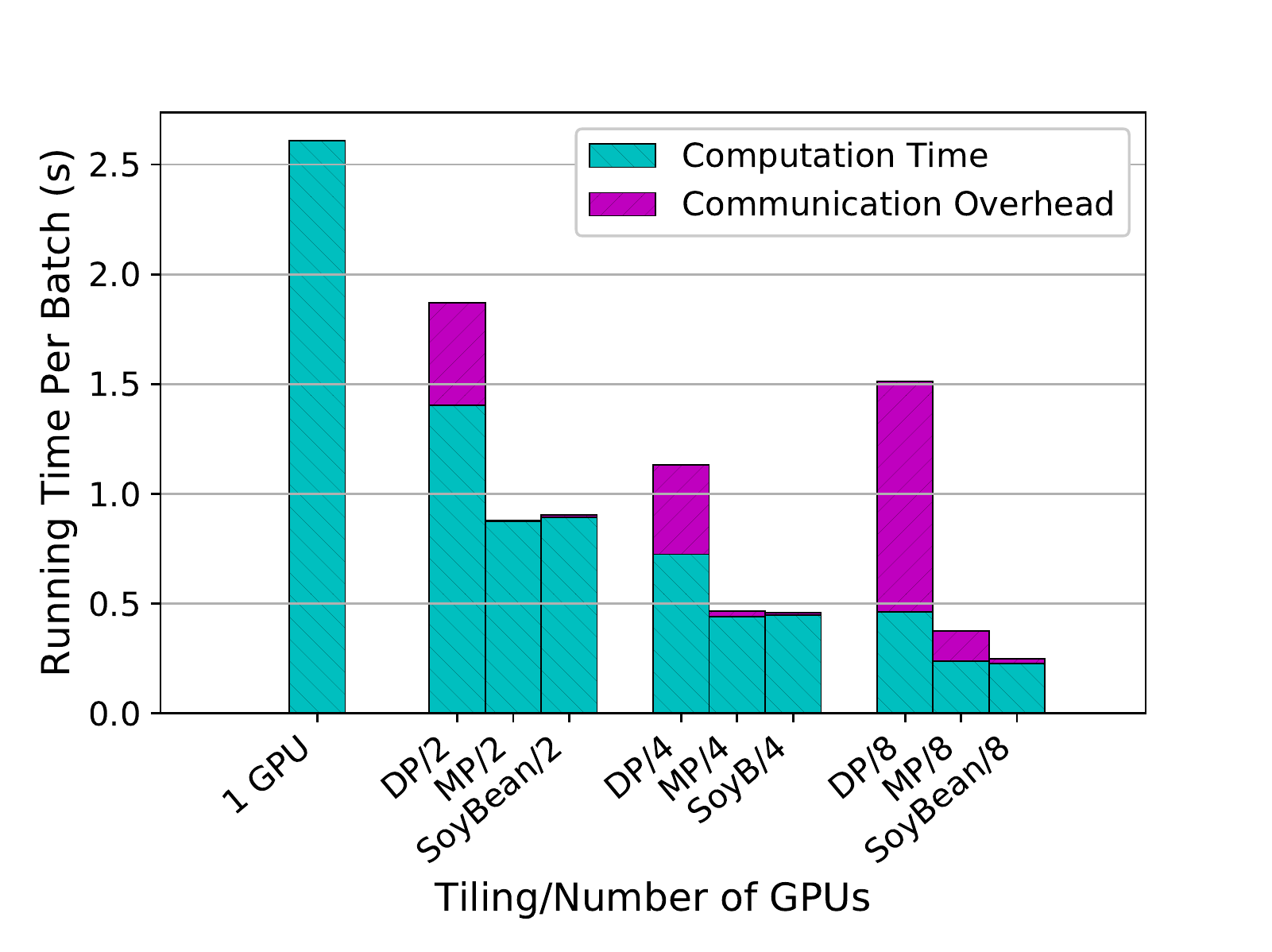}}
		\caption{\small{Runtime comparison of a 4-layer MLP for DP, MP, and \name
				with different batch sizes and hidden sizes.
				~\label{fig:mlp}}}
	\end{center}
\end{figure*}

In this section, we examine \name's performance.
Specifically, we want to answer following questions:
\begin{enumerate}
	\item Is communication really a bottleneck when using data parallelism with
	small batch sizes?
	\item Can \name reduce overall runtime?
	\item How well does \name accelerate modern DNNs?
\end{enumerate}

\subsection{Experimental Setup}
We evaluated \name on Amazon's EC2 cluster. We used a p2.8xlarge instance, with 480GB of memory and 32 virtual CPUs as well as 8 NVIDIA GK210 GPUs on the instance. Each of the GPUs has 12GB of memory; they are connected by PCI-e, with a maximum peer-to-peer bi-directional bandwidth of 20GB/s.

\subsection{Communication Overhead Evaluation}
\label{sec:microbenchmark}
In this evaluation, we want to know if communication overhead accounts for
a large percentage of the overall runtime when the batch size is relatively small
or when the weight size is large.
We first tested the runtime for data parallelism (DP), model parallelism (MP), and \name
with different parameters, i.e., batch size and weight size.
To see the ratio of the communication overhead to runtime, we also
modified \mxnet's backend to skip any communication, and re-ran our experiments.
Thus, the runtime measured by the modified backend is solely due to computation.
We can then determine the communication overhead by subtracting the computation time from
the original runtime. Note that we report communication \emph{overhead} instead of
communication time because communication can be overlapped with computation.
As a result, the communication overhead is strictly smaller than the communication time.

Figure~\ref{fig:mlp}(a) and Figure~\ref{fig:mlp}(b) show the runtime and
communication overhead of a 4-layer MLP for different tilings on different numbers of GPUs.
Both tests have the same weight size, 8192$\times$8192, but different batch sizes, 512 and 2048,
respectively.
In Figure~\ref{fig:mlp}(a), the communication overhead for data parallelism increases as the
number of GPUs increases. This is because a GPU needs to send its data to more GPUs when the
total number of GPUs increases. However, the aggregate communication throughput is limited by
contention on shared PCI-e resources, so it cannot increase linearly with several simultaneous
peer-to-peer connections.
As a result, increased GPU-to-GPU communication is slower, even below the PCI-e bandwidth limit.

In these two figures, data parallelism performs sq worse than MP and \name. That's
because both communication overhead and computation time in data parallelism are larger than
MP and \name. We will show why different tilings may result in different computation
time in Section~\ref{sec:shape}. However, even if the computation time is the same as with MP
and/or \name, data parallelism is still slower.
The key reason is the communication overhead is huge, \mytilde5$\times$ longer than the 
computation time on 8 GPUs with batch size 512, and \mytilde2.5$\times$ longer on 8 GPUs
with batch size 2048.

Figure~\ref{fig:mlp}(c) uses the same batch size as Figure~\ref{fig:mlp}(b) but 
with larger weight size, 12288 (12K). 
Unlike changing the batch size, which significantly affects the ratio of communication overhead
to total runtime, changing the weight size from 8K to 12K has little effect on the ratio.
This is because when changing the weight size, both communication and computation increase for all
parallelism schemes.

The results show that to achieve good performance, one should not use data parallelism
when the batch size is small, due to the communication overhead. 
As the batch size increases, the percentage of the runtime consumed by communication overhead becomes
smaller for data parallelism, and a crossover point will be reached when its runtime is lower than
model parallelism on the same model and batch size.
Because \name can use hybrid tilings, it always achieves optimally low communication overhead.

\begin{figure*}
	\centering
	\subfloat[Image Size = 6x6, Filter Size = 2048]{%
		\includegraphics[width=.45\linewidth]{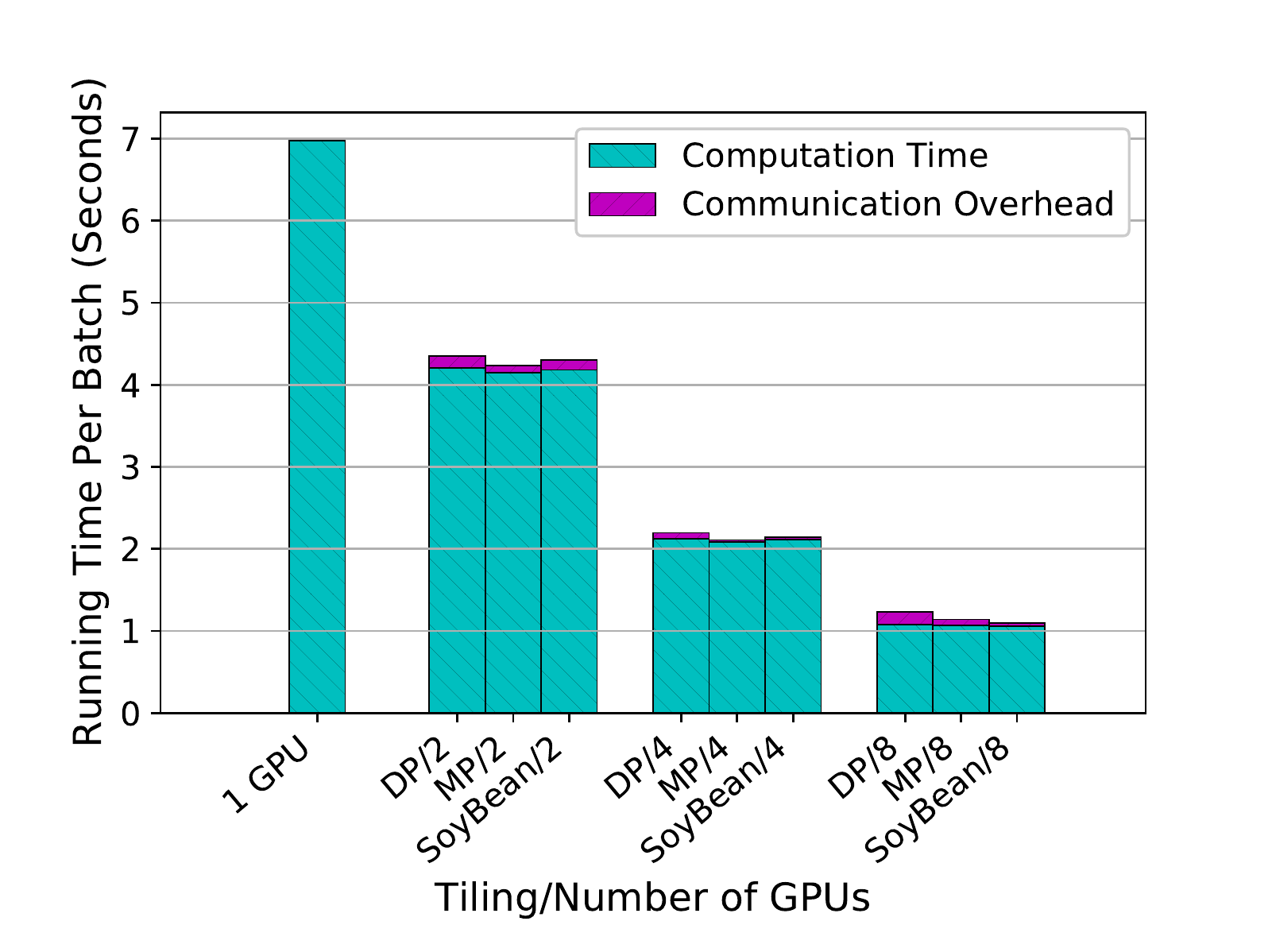}}
	\subfloat[Image Size = 24x24, Filter Size = 512]{%
		\includegraphics[width=.45\linewidth]{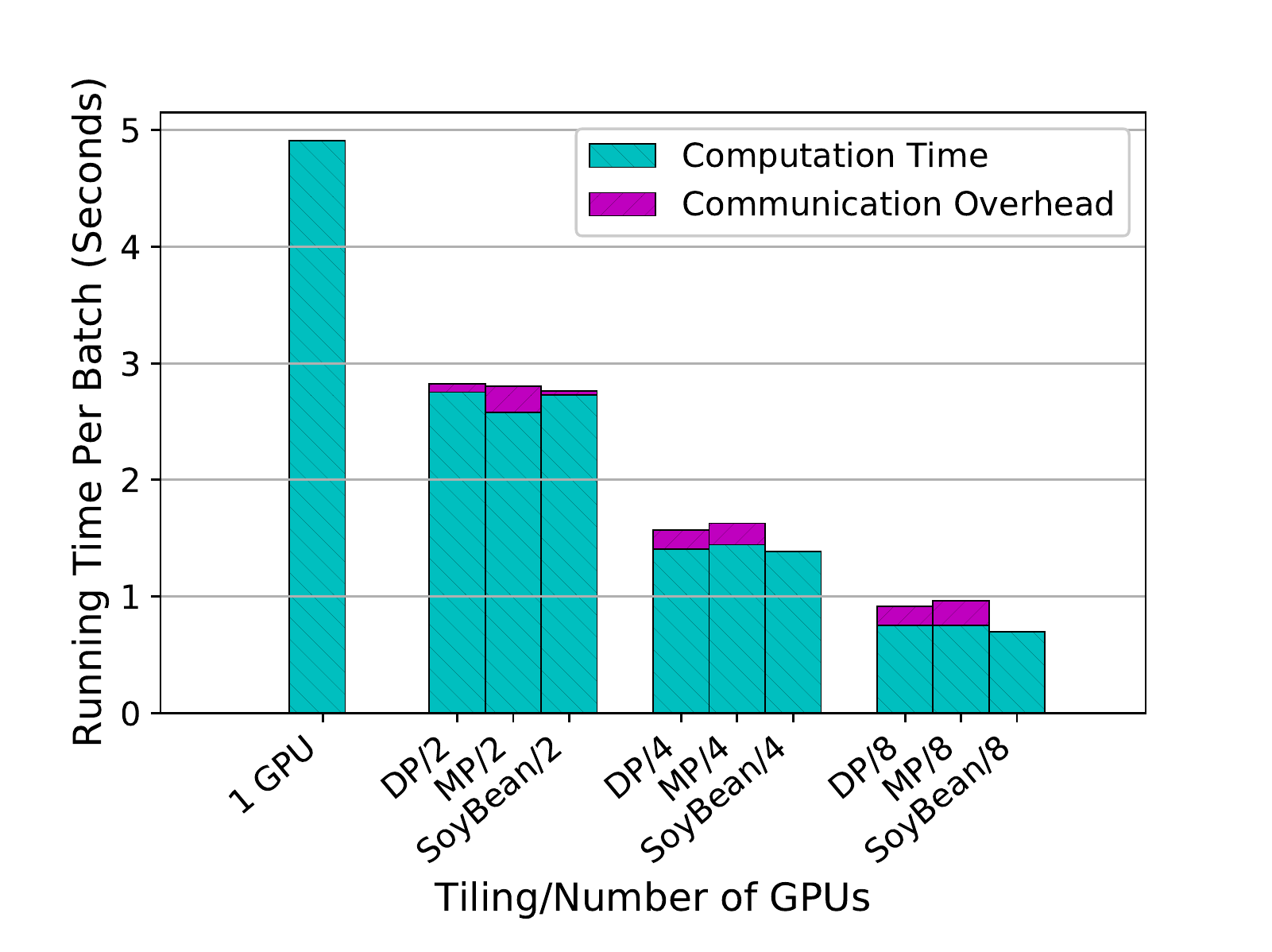}}
	\caption{\small{Runtime comparison of training a 5-layer convolutional neural network
			using DP, MP, and \name. The batch size is 256.
			~\label{fig:conv}}}
\end{figure*}

We also performed the same evaluation for convolution neural networks.
Figure~\ref{fig:conv}(a) shows results from training a CNN with small (6px$\times$6px) images
with a large filter size (2K), while Figure~\ref{fig:conv}(b) shows results from training with larger
(24px$\times$24px) images with a small filter size (512).
We fixed the batch size to 256 for both experiments. 
The results show that with the larger image size,
data parallelism has better performance than model parallelism.
\name still outperforms both partitioning schemes due to its ability to cut in different dimensions.

\subsection{Can Shape Affect Computation Performance?}
\label{sec:shape}
\begin{table}[h]
	\centering
	\small
	\begin{tabular}{c|c|c}
		\hline\hline
		Batch Size & Single GPU & \begin{tabular}{c} Single GPU \\ w/ \name tilings \end{tabular}\\
		\hline\hline
		512 & 0.31s & 0.19s \\
		\hline
		1024 & 0.56s & 0.39s \\
		\hline
		2048 & 1.13s & 0.73s \\
		\hline
	\end{tabular}
	\caption{\small{Runtime per batch comparison for a 4-layers MLP network
			between single GPU and single GPU with \name partitions.
			The weight size is fixed to 8K$\times$8K.}}
	\vspace{0.5em}
	\label{fig:single_partition}
\end{table}
Figure~\ref{fig:mlp} shows that the computation time varies with the tiling approach
chosen. Regardless of the tiling, however, the total amount of computed
data is the same. Therefore, we suspected that the shapes of matrices might affect the
computation performance. To validate this assumption, we partitioned the input matrices into
several sub-matrices by using \name, but put all of them on single GPU. We achieved better
performance with this approach than with using the un-cut matrices to do the same
computation on a single GPU, as shown in table~\ref{fig:single_partition}.
We believe the difference is related
to how CUDA~\cite{cuda} chooses different algorithms according to the matrix shapes.

\subsection{Scalability}
\begin{figure*}
	\centering
	\subfloat[AlexNet throughput speedup.]{%
		\includegraphics[width=0.45\textwidth]{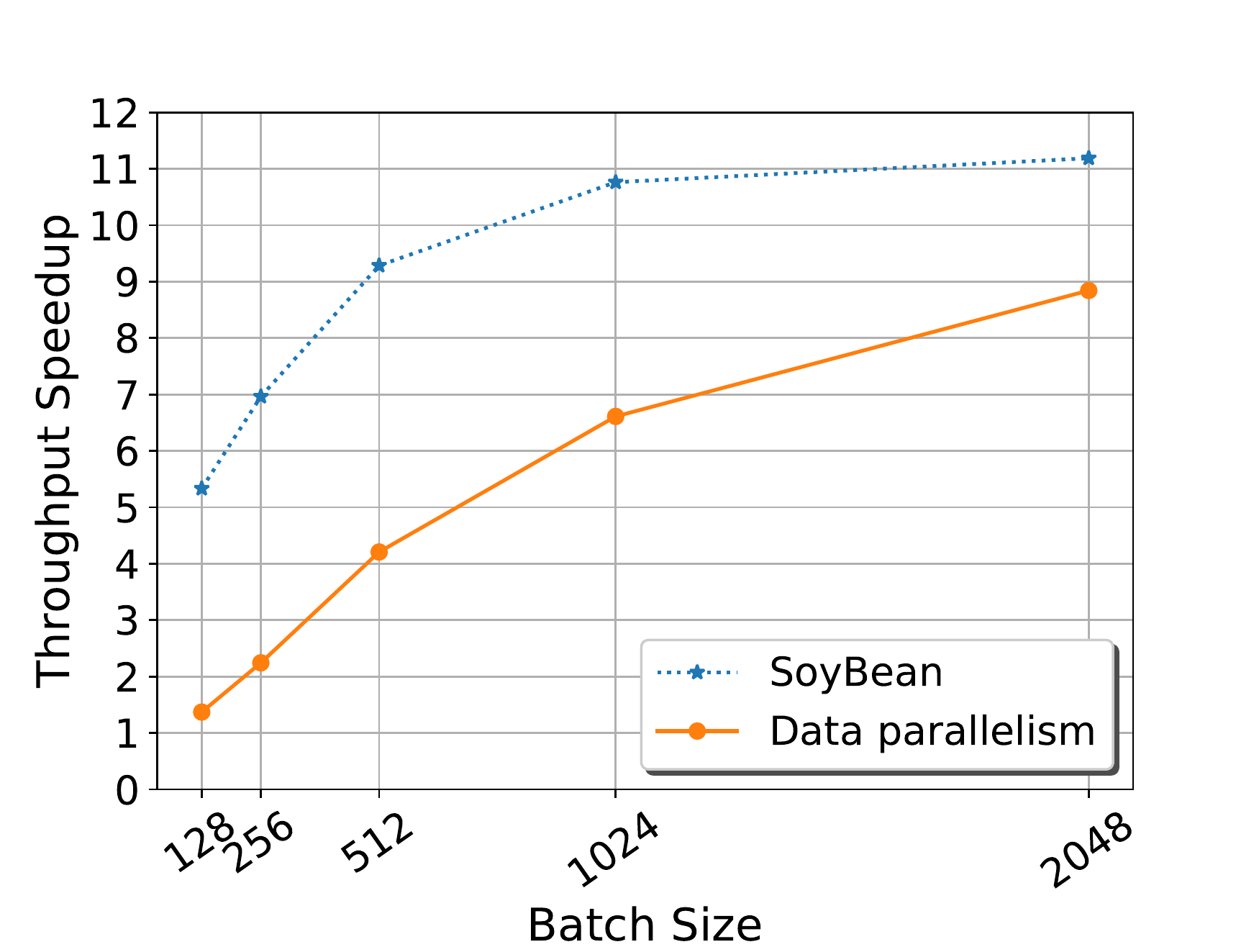}}
	\subfloat[VGG throughput speedup.]{%
		\includegraphics[width=0.45\textwidth]{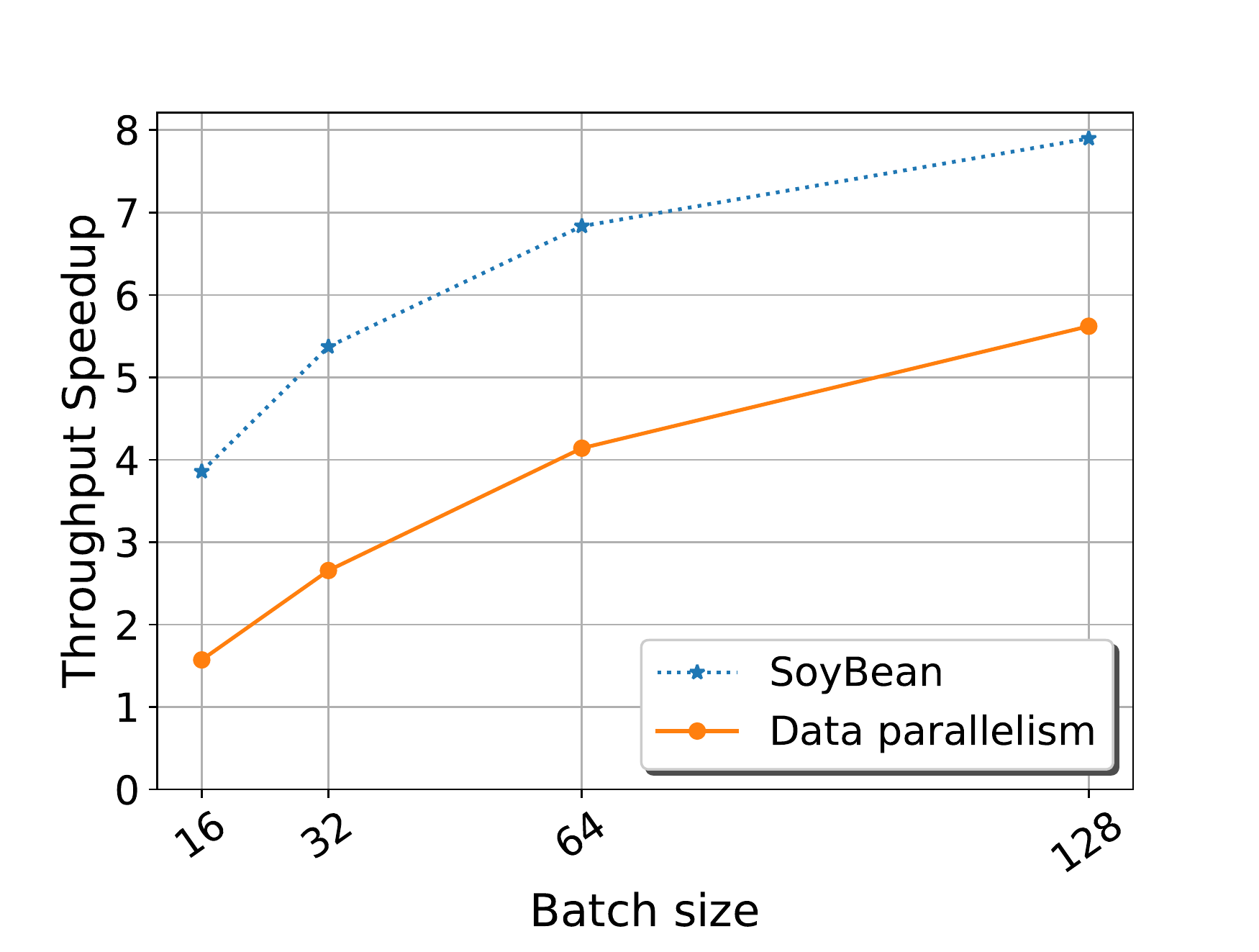}}
	\caption{\small{Throughtput comparison of \name and data parallelism on 8 GPUs.
			~\label{fig:speedup}}}
\end{figure*}

Our goal is to achieve good performance regardless of the model and batch sizes; particularly
difficult is good scalability with smaller batch sizes, which data parallelism does not achieve.
In this section, we evaluate \name's scalability with two popular image recognition neural networks,
AlexNet and VGG on 8 GPUs. We first trained each network on a single GPU to determine the
maximum throughput (images/second). Using this throughput as baseline, we then calculated
the speedup for different batch sizes.

Figure~\ref{fig:speedup}(a) shows the results with AlexNet. \name achieves a greater
than $7\times$ speedup with a batch size of 256, while data parallelism requires increasing
the batch size to more than 1K to achieve the same speedup.
AlexNet consists of many convolution layers followed by
fully connected layers. As discussed in Section~\ref{sec:microbenchmark},
data parallelism needs a large batch size to achieve good scalability for fully connected
layers. Moreover, Section~\ref{sec:microbenchmark} also shows that \name can always
achieve similar or better performance for convolution layers than data parallelism.
As a result, \name performs much better than data parallelism.
VGG has similar structure to AlexNet but with more layers. Therefore, \name
can still achieve better scalability than data parallelism as shown in
Figure~\ref{fig:speedup}(b). One may notice that in Figure~\ref{fig:speedup}(a),
\name can achieve superlinear speedup. This is because some matrix shapes fall into categories
that have better computation performance after partitioning, as
discussed in Section~\ref{sec:shape}.


\section{Related Work}
\label{sec:related}
How to distribute arbitrary data easily for users while achieving
high-performance computation at the same time has been a popular research topic.
However, it is difficult to design systems that automatically optimizes locality
without knowledge of the underlying data and processing.
Relatedly, distributed array programs are increasingly
important due to the emergence of machine learning and deep learning.
As a result, significant effort has been expended in optimizing distributed array frameworks.

\noindent{\bf Deep learning systems. }
Many frameworks, such as Tensorflow~\cite{tensorflow}, MXNet~\cite{mxnet},
PyTorch~\cite{pytorch}, Theano~\cite{theano} and Caffe2~\cite{caffe2} have
been proposed to facilitate developing new neural network models.
These distributed array frameworks emphasize deep learning and machine learning applications.
Besides the common array operations,
they also provide many functionalities for neural networks such as automatic
differentiation (backpropagation).

Though these frameworks allow users to develop models with both data parallelism
and model parallelism, their interfaces are not friendly for utilizing model parallelism.
\name not only simplifies exploit model parallelism, but can automatically
choose correct combinations of various parallelisms to reduce communication.
Moreover, because of the similar design approach, performing optimizations over
data-flow graphs, \name can be implemented in these frameworks as long as
they allow customized optimizations. 

\noindent{\bf General distributed programming frameworks.}
Most distributed frameworks target primitives for key-value collections (e.g.
MapReduce~\cite{mapreduce}, Dryad~\cite{dryad}, Piccolo~\cite{piccolo},
Spark~\cite{spark}, Ciel~\cite{ciel}, Dandelion~\cite{dandelion} and
Naiad~\cite{naiad}). Some provide graph-centric primitives (e.g.
GraphLab~\cite{graphlab} and Pregel~\cite{pregel}).
It is possible to implement a deep learning framework backend by augmenting
an in-memory framework, such as Spark or Piccolo. When doing so, \name can
be applied to optimize the high-level data-flow graph before lowering the
graph to the underlying framework.

\noindent{\bf Distributed array frameworks.}
Relational queries are a natural layer on top of key-value centric distributed
execution frameworks, as seen in systems like DryadLINQ~\cite{dryadlinq},
Shark~\cite{shark}, Dandelion~\cite{dandelion} and Dremel~\cite{dremel}.
Several attempts have been made to build array interfaces on these.
MadLINQ~\cite{madlinq} adds support for distributed arrays and array-style
computation to the dataflow model of DryadLINQ~\cite{dryadlinq}.
SciHadoop~\cite{SciHadoop} is a plug-in for Hadoop to process array-formatted data.
Google's R extensions~\cite{googler}, Presto~\cite{presto} and
SparkR~\cite{sparkr} extend the R language to support distributed arrays.
Julia~\cite{Julia} is a newly developed dynamic language designed for
high performance and scientific computing. Julia provides primitives for users to
parallel loops and distribute arrays.
Theoretically, one can use these extensions and languages to implement any
neural network models. However, users have to manually deal with all optimizations
provided by deep learning frameworks, like backpropagation (and with \name, tiling).

Spartan~\cite{spartan} and Kasen~\cite{kasen} are two distributed array frameworks
that perform automatic tiling on general array programs.
Both systems provide primitives operating directly on arrays and thus expose the
data access pattern of array operations that can be used for tiling optimization.
Spartan proves that under such setting, automatic tiling is an NP-hard problem and
provides only heuristic algorithms to find the best tiling.
By contrast, \name specifically targets deep learning systems and observes 
many fixed graph structures in neural network models. These fixed graph structures allow 
\name to be able to find the optimal solutions in most cases. 
Moreover, \name doesn't rely on specific
primitives, but directly optimizes array operations.

\noindent {\bf Distributed array libraries.}
Optimized, distributed linear algebra libraries, like LAPACK~\cite{lapack},
ScaLAPACK~\cite{scalapack}, Elemental~\cite{Elemental} Global
Arrays Toolkit~\cite{globalarrays} and Petsc~\cite{petsc-user-ref, petsc-efficient}
expose APIs specifically designed for large matrix operations.  They focus
on providing highly optimized implementations of specific operations. However,
their speed depends on correct partitioning of
arrays and their programming models are difficult for use in deep learning.

\noindent{\bf Compiler-assisted data distribution. }
Prior work in this space proposes static, compile-time techniques for analysis.
The first set of techniques focuses on partitioning~\cite{hudak1990compiler};
the second on data co-location~\cite{knobe1990data, philippsen1995automatic, milosavljevic1999automatic}.
Prior work also has examined nested loops with affine array subscript patterns,
using different structures (vector~\cite{hudak1990compiler},
matrix~\cite{ramanujam1991compile} or reference~\cite{ju1992reduction}) to
model memory access patterns or polyhedral model~\cite{lu2009data} to perform localization analysis.
Since static analysis deals poorly with ambiguities in source
code~\cite{softwareengineering}, recent work proposes profile-guided methods~\cite{chu2007data}
and memory-tracing~\cite{park2014trace} to capture memory access
patterns. Simpler approaches focus on examining stencil code~\cite{park2014trace,
	he2011automatic,
	hernandez2013open64, kessler1996pattern, henretty2011data}.

Access patterns can be used to find a distribution of data that minimizes
communication cost~\cite{hudak1990compiler,ramanujam1989methodology, bau1995solving,
	d1989partitioning, huang1993communication}.  All approaches construct a weighted graph
that captures possible layouts.  Although searching the optimal
solution is NP-Complete~\cite{kennedy1998automatic, kremer1993np, li1990index, li1991data},
heuristics perform well in practice~\cite{li1991data, philippsen1995automatic}.
DMLL~\cite{dmll} aims to extend a data-parallel programming mode to heterogeneous hardware.
To achieve the goal, DMLL propose a new intermedia language based on common parallel patterns
to partition data and distributed across the underlying heterogeneous distributed hardware.

\name borrows many ideas from these works, such as constructing a weighted graph.
However, unlike prior work that requires language-specific extensions and/or modification
to capture parallel access patterns, \name is designed to utilize the parallelism exposed
by the underlying data-flow graphs, and can be quickly integrated with modern
deep learning frameworks.

\section{Conclusion}
\label{sec:concl}
Deep learning systems have become indispensible in many fields, but as their complexities grow, DNN training time is becoming increasingly intractable. We demonstrate effective tiling in \name that can achieve performance as good as or better than the best of data parallelism and model parallelism for many types of DNNs on multiple GPUs. With the speed and simplicity provided by \name's backend, users can train networks using frontends like \tf and \mxnet quickly and easily, making DNNs useful for a larger audience.

%
%
%

\bibliographystyle{abbrvnat}

%
%
%

\end{document}